# Detection of Obstructive Sleep Apnoea Using Features Extracted from Segmented Time-Series ECG Signals Using a One Dimensional Convolutional Neural Network

Steven Thompson
*Computer Science*
*Liverpool John Moores University*
Liverpool, Merseyside
S.R.Thompson@LJMU.AC.UK

Paul Fergus
*Computer Science*
*Liverpool John Moores University*
Liverpool, Merseyde
P.Fergus@LJMU.AC.UK

Carl Chalmers
*Computer Science*
*Liverpool John Moores University*
Liverpool, Merseyside
C.Chalmers@LJMU.AC.UK

Denis Reilly
*Computer Science*
*Liverpool John Moores University*
Liverpool, Merseyside
D.Reilly@LJMU.AC.UK

*Abstract*—The study in this paper presents a one-dimensional convolutional neural network (1DCNN) model, designed for the automated detection of obstructive Sleep Apnoea (OSA) captured from single-channel electrocardiogram (ECG) signals. The system provides mechanisms in clinical practice that help diagnose patients suffering with OSA. Using the state-of-the-art in 1DCNNs, a model is constructed using convolutional, max pooling layers and a fully connected Multilayer Perceptron (MLP) consisting of a hidden layer and SoftMax output for classification. The 1DCNN extracts prominent features, which are used to train an MLP. The model is trained using segmented ECG signals grouped into 5 unique datasets of set window sizes. 35 ECG signal recordings were selected from an annotated database containing 70 night-time ECG recordings. (Group A – a01 to a20 (Apnoea breathing), Group B – b01 to b05 (moderate), and Group C – c01 to c10 (normal). A total of 6514 minutes of Apnoea was recorded. Evaluation of the model is performed using a set of standard metrics which show the proposed model achieves high classification results in both training and validation using our windowing strategy, particularly W=500 (Sensitivity=0.9705, Specificity=0.9725, F1_Score=0.9717, Kappa_Score=0.9430, Log_Loss=0.0836, ROCAUC=0.9945). This demonstrates the model can identify the presence of Apnoea with a high degree of accuracy.

## I. Introduction

Obstructive Sleep Apnoea (OSA), is a sleep disorder that interrupts the natural rhythm of a person's breathing whilst they are sleeping. In the International Classification of Sleep Disorders Third Edition (ICSD-3) report, published in April 2014, by the international classification of sleep disorders, OSA is classified as the most common subtype of breathing disorder of sleep (SDB) and is characterised by episodes of complete or partial upper airway obstruction during sleep. The symptoms of Sleep Apnoea include chronic snoring, insomnia, gasping and breath holding, unrefreshing sleep, and daytime sleepiness [1]. OSA can affect anyone regardless of age or gender, however, most studies show the condition to be more prevalent amongst 30 to 60 year olds.[2][3]. The Apnoea–Hypopnoea Index (AHI) is used to indicate the severity of OSA with an AHI value <5 classed as normal. Estimates have shown that OSA affects 20% of the general population, where AHI is ≥5 [4]. However, despite the prevalence of OSA, the vast majority of OSA sufferers go undiagnosed [2][5].

Current diagnostic techniques for OSA can be expensive, cumbersome, complex and lengthy, meaning sufferers do not receive the required treatment and therapy needed. It has been suggested that over 80% of patients remain incorrectly diagnosed [5][6]. Consequently, OSA represents a major public health concern and left untreated can lead to numerous negative health-related consequences and in some cases mortality [7][8]. OSD results in a lack of sleep and/or poor sleep quality, which can affect an individual's function and decision-making capabilities. This can often lead to accidents at home, their ability to drive and in accidents in the workplace [9]. Globally, the direct and indirect costs of OSA, such as health care costs, accidents, decreased productivity and sickness reaches billions annually [6].

Diagnosing OSA is determined through consultation with a physician or sleep specialist. A physical examination is performed to consider the blood pressure, body mass index (BMI) and neck measurements of the patient. This is often followed up by a detailed discussion to gather sleep information, typically achieved using a sleep log or sleep diary to record sleep times, nightly bedtimes, time to fall asleep, morning arising-times, wake-up-time duration and number, additional nap times and any episodes of tiredness/sleepiness throughout the day [10]. Other mechanisms include self-assessment questionnaires, such as the Epworth Sleepiness Scale (ESS) [11], Berlin [12] and STOP-Bang Questionnaires [13].

A more precise diagnosis and information gathering process can be performed through non-intrusive sleep studies, also known as Polysomnography (PSG). PSGs are the best approach when identifying incidences of OSA and are the preferred method for clinicians [10]. It involves the recording and analysis of multiple physiological variables during sleep using body worn sensors to record electroencephalogram (EEG), electrooculography (EOG), electromyography (EMG), electrocardiogram/ electrocardiography (ECG or EKG), nasal cannulas, pulse oximeters and respiratory belts. Patients will generally sleep overnight in a PSG sleep centre attached to as many as 16 separate physiological sensor channels and multiple devices to monitor stages of sleep, measure oxygen levels, body movements, heart rate and breathing patterns, to provide a comprehensive analysis [14]. However, there are several major problems with this type of diagnostic testing. These include a lack of available PSG sleep centres and equipment, and high costs and employment of sleep technicians to monitor a person's sleep [15]. Furthermore, it is often an inconvenience for patients to attend and actually sleep in sleep laboratories, particularly when testing on children [15].

To combat these issues alternative OSA diagnostic methods have been proposed. One example is the Home Sleep Apnoea Testing (HSAT) kit, known in Europe as polygraphy kits. HSATs are lightweight, portable and wearable devices that use far less physiological sensors than standard PSGs [16]. With better accessibility, reduced waiting lists and lower overall costs, HSATs are used for first line diagnosis of OSA and a substitute to PSG sleep centres. However, their use as stand-alone diagnostics in routine clinical practice is yet to yield any convincing results [17]. This is primarily because HSATs find it difficult to compute the Respiratory Event Index (REI), since it is calculated against recoding-time instead of sleep-time and this ultimately misrepresents AHI assessments [18].

The use of Data Science has also featured in several studies to provide a data-driven methodology in OSA detection. Harnessing the power of advanced machine learning algorithms with clinical expertise, it is now possible to produce better diagnostic results using single-channel physiological signals. This method dramatically reduces the amount of required equipment, time and costs, thus overcoming many of the PSG shortcomings and provides a platform for novel studies and proposals which included the use of ECG [19], EEG [20], Nocturnal Oximetry recordings [21] Respiratory sensors [22][23] and Snoring audio segments [24][25][26].

## II. Materials and Methods – Data acquisition, Subject Information and Pre-processing

### A. Apnoea-ECG Database

Penzel et al. [27] conducted a comprehensive study between 1993 and 1995 to investigate and record the effect of OSA on arterial blood pressure in subjects with moderate and severe Sleep Apnoea. A second study undertaken between 1998 and 1999, was to create a normative set of sleep recordings, with the main research focused on multi-channel EEG recordings performed on healthy volunteers and patients suffering with Sleep Apnoea. Here, they have combined records and ECG recordings from both of these studies to create a single database (Apnea-ECG database), publicly available via Physionet.

The Apnoea-ECG database contains the records of 70 patients (subjects). The dataset comprises a mixture of male and female subjects with ages ranging from 27 to 63 years (mean 45yrs). Body mass index (BMI) recordings vary between 19.2 and 45.33kg. (mean: 28.01 ± 6.49 kg.) and body weights range between 53 to 135 kg (mean: 86.3 ± 22.2 kg). Only 35 records have associated annotations - (a01 to a20 (20 ECG signals), b01 to b05 (5 ECG Signals), and c01 to c10(10 EGC signals). The 35 non-annotated records were removed. Each of the ECG annotated recordings vary in length from approx. 7hrs to 10 hours. The three groups within the recordings are defined by the AHI index. The AHI index aross these groups varied between 5 and 82 events per hour. Group A (Apnoea-Set): each subject has over 100 minutes of Apnoea; Group B (Borderline-Set): this group has between 5 to 99 minutes of Apnoea; and Group C (Normal-Set): each subject in this group has between 0 to 3 minutes of Apnoea. A total of 17,125 minutes (or 285hrs 25mins) sleep-time was recorded; of this, 6,514 minutes (or 108hrs 34mins) was scored as Apnoea and 10,611 minutes (176hrs 51mins) was scored as Non-Apnoea. Table I provides a summary for the group distributions.

TABLE I.

| Subject Recordings | EGC Files | Group Type | Apnoea Events (Mins) | Non-Apnoea (Mins) |
|---|---|---|---|---|
| A01 – A20 | 20 | Apnoea-Set | 6250 | 3811 |
| B01 – B05 | 5 | Borderline-Set | 252 | 2060 |
| C01 – C10 | 10 | Normal-Set | 12 | 4740 |
| | | | 6514 | 10611 |

### B. Annotations

The recordings were labelled by an expert scorer for Sleep Apnoea events. Each observation is labelled for each 60-second block indicating the presence (or absence) of events in that segment. ECG signals where sampled at 100 Hz. The resolution of the signal is 12-bit. So, each ECG signal segment is 60s or 6000 samples long. Each "A" annotation indicates that Apnoea was in progress at the beginning of the associated minute; each "N" annotation indicates that Apnoea was not in progress at the beginning of the associated minute. The Apnoea index (AI) is the number of Apnoeas observed per hour, and the HI is the number of hypopneas observed per hour. The AHI is defined as the sum of AI and HI.

### C. Segmentation

Cross-referencing each of the ECG recordings with their associated annotation file. All Apnoea events and Non-Apnoea events from each recording was separated into two groups. For example, Table II shows how each ECG recording from the 3 groups (A, B and C) were separated and placed into either the Apnoea and the Non-Apnoea group. Performing this procedure on each individual signal recording, resulted in 650 separate event files; 314 Apnoea and 336 Non-Apnoea.

TABLE II.

| 35 ECG Recording | Apnoea events (segmented files) | Non-Apnoea events (segmented files) |
|---|---|---|
| A01 | 3 | 3 |
| A02 | 11 | 11 |
| A03 | 11 | 11 |
| A04 | 3 | 3 |
| A05 | 15 | 16 |
| A06 | 10 | 11 |
| A07 | 23 | 23 |
| A08 | 32 | 33 |
| A09 | 14 | 14 |
| A10 | 18 | 18 |
| A11 | 7 | 7 |
| A12 | 7 | 7 |
| A13 | 20 | 20 |
| A14 | 8 | 9 |
| A15 | 16 | 17 |
| A16 | 13 | 14 |
| A17 | 14 | 15 |
| A18 | 6 | 6 |
| A19 | 16 | 16 |
| A20 | 16 | 17 |
| B01 | 6 | 7 |
| B02 | 13 | 14 |
| B03 | 11 | 12 |
| B04 | 3 | 4 |
| B05 | 7 | 7 |
| C01 | 0 | 1 |
| C02 | 1 | 2 |
| C03 | 0 | 1 |
| C04 | 0 | 1 |
| C05 | 2 | 3 |
| C06 | 1 | 2 |
| C07 | 4 | 5 |
| C08 | 0 | 1 |
| C09 | 2 | 3 |
| C10 | 1 | 2 |
| Total Segmented | 314 | 336 |

### D. Dataset formation

The newly segmented files (Table II) were used to form 5 balanced datasets of different window sizes, each containing equal amounts of Apnoea and Non-Apnoea events, as seen in Table IV. All Apnoea events are labelled as "1" and all Non-Apnoea events as "0".

Both Apnoea and Non-Apnoea types are combined resulting in 70 merged files (35 Apnoea files and 35 Non-Apnoea files), Table III.

TABLE III.

| Subject Recordings | Apnoea events (merged files) | Non-Apnoea events (merged files) |
|---|---|---|
| A01 – A20 | 20 | 20 |
| B01 – B05 | 5 | 5 |
| C01 – C10 | 10 | 10 |
| Total | 35 | 35 |

### E. Windowing Strategy

The 70 signal recordings are individually shaped into the specific window sizes of, 500, 1,000, 1,500, 2,000 and 2,500 respectively. Through controlled techniques of squaring and merging, each recording was stacked until 5 datasets with balanced classes were formed. W=500, W=1000, W1500, W=2000 and W=2500, as seen in Table IV.

TABLE IV.

| Dataset | Window Size | Apnoea | Non-Apnoea |
|---|---|---|---|
| W=500 | 500 columns | 35 files | 35 files |
| W=1000 | 1,000 columns | 35 files | 35 files |
| W=1500 | 1,500 columns | 35 files | 35 files |
| W=2000 | 2,000 columns | 35 files | 35 files |
| W=2500 | 2,500 columns | 35 files | 35 files |

Table V shows the structure and dimensions for each of the 5 datasets. This includes the amount of columns (window size), rows (samples) and signal readings, which is approx. 75,000,000 per dataset. Each dataset is built up starting with Non-Apnoea samples taking up the bottom half of the dataset and represented with the number '0' at the end of each sample. The top half is completed with Apnoea samples and represented at the end of each sample with the number 1.

TABLE V.

| Dataset | Columns | Row Samples | Signal bits |
|---|---|---|---|
| W=500 | 500 | 150,384 | 75,192,000 |
| W=1000 | 1,000 | 75,190 | 75,190,000 |
| W=1500 | 1,500 | 50,126 | 75,189,000 |
| W=2000 | 2,000 | 37,592 | 75,184,000 |
| W=2500 | 2,500 | 30,060 | 75,150,000 |

### F. One Dimensional Convolutional Neural Network Model

CNNs have become an established tool in deep learning research. Two and three-dimensional models, are used for complex tasks, such as image processing and shape recognition. Image processing typically use three dimensions (one each for RGB vectors). In this study however, time-series signal-data is used, which only requires a 1-dimentional CNN.

The 1DCNN architecture in this study (Fig.1) is constructed with several layers which include; 1 Convolutional layer, 1 Max Pooling layer and a Fully Connected Multilayer Perceptron (MLP) consisting of 1 Hidden layer and a Softmax output layer.

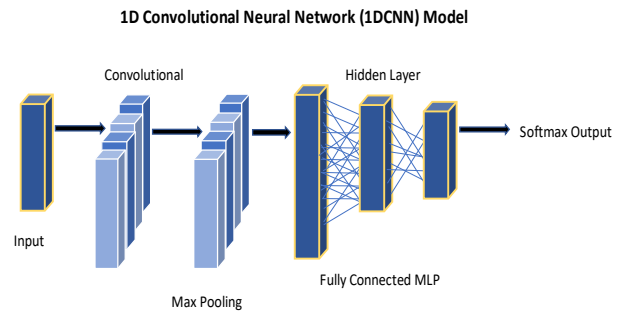

Fig. 1. Architecture of our One-Dimensional Convolutional Neural Network.

Single-channel ECG-signal data is presented at the input layer of the IDCNN. Features are extracted from pre-configured input vectors using a single convolution layer were the data is processed through a series of filters. The filters convolve the data, each time extracting the most relevant and prominent features to build a map of activations (feature map), whilst automatically learning each of the filter parameters. To assist the convolutional process, a max pooling layer is introduced to reduce the selected elements on each of the created feature maps, while retaining the most prominent of these elements. The overall aim is to simplify the convolutional layer output, meaning less computation, reducing overfitting and improving the model's performance. The data is then passed to a Fully Connected Multilayer Perceptron (MLP) which controls learning and reduces errors. The MLP consists of 3 layers; input layer, which receives our signal data; a hidden layer containing ReLU activation function, where the main computation is performed; and an output layer for softmax classification. Using backpropagation, the data is passed back and forth through these layers to continually train the network and minimise errors, whilst an implemented ADAM optimizer helps to reduce the difference between the predicted output and actual output.

## G. Performance Metrics

This section provides a brief description of the performance metrics used throughout our testing phase. These metrics provide an indication of how well our model is performing using specific parameters and configurations.

### 1) Sensitivity (Recall) and Specificity

Sensitivity (Recall) and specificity are two common performance metrics which measure the classification of an instant into two groups. Sensitivity measures the true positive rate (it measures the number of actual positives that are correctly identified). Specificity measures the true negative rate (measures the proportion of actual negatives that are correctly identified)

### 2) ROC AUC

The area under the ROC Curve (AUC) is one of the most common and important measurement tools used for diagnostic accuracy. It gives a graphical representation of how confident a model is at distinguishing between two classes. This is presented by ranking the two separate classes on a scale of 0 to 1. Generally, the higher the AUC the better the model is at prediction and class separability.

$$\text{True Positive Rate (TPR)} = \frac{TP}{TP + FN} \quad (1)$$

$$\text{False Positive Rate (FPR)} = \frac{FP}{FP + TN}$$

### 3) F1_Score

The F1 score is a binary classifier measurement of test accuracy. It performs this by calculating the mean of precision and recall. The higher the recall and precision results, the greater the F1 score.

$$F1 = 2 * \frac{precision * recall}{precision + recall} \quad (2)$$

### 4) Kappa_Score

Kappa score represents the level of agreement between two variables on a classification problem. The Kappa statistic is frequently used to test inter-rater reliability.

$$k = \frac{\Pr(a) - \Pr(e)}{1 - \Pr(e)} \quad (3)$$

### 5) Log_loss

The function of Log Loss is to measure the accuracy of a classifier. This is achieved through the confidence of the classification compared to the actual result. The greater the correctly predicted probability, the smaller the logloss, which in turn means a better accuracy.

$$logloss = -\frac{1}{N}\sum_{i=1}^{N}[y_i \log(\hat{y}_i) + (1 - y_i)\log(1 - \hat{y}_i)] \quad (4)$$

### 6) Loss & Accuracy

These two functions are used on the training set and cross-validation technique when combined with the Validation Loss and Validation Accuracy. The Loss function optimises the model and the Accuracy function then measures the performance of the model. These calculations are performed after every batch and give an overall measure of how the model is progressing in terms of training.

### 7) Validation Loss & Validation Accuracy

Validation loss and Validation Accuracy is the same metric as loss and accuracy, but they are not used to update the weights. It is calculated in the same way, but this calculation is used to evaluate the quality of the model by evaluating its performance after every epoch.

## III. EXPERIMENTS AND RESULTS

In this section we look to fully evaluate the overall efficiency and performance of our proposed 1DCNN when performing feature extraction and classification tasks, using balanced datasets that incorporates a windowing strategy. A total of 5 experiments where conducted, one for each window size, W=500, W=1000, W=1500, W=2000 and W=2500. These experiments were measured and evaluated using a selection of standard metrics. The data split used for each experiment was 72% training, 20% testing and 8% for validation. Tables V through to IX shows each model's optimum configuration (inputs) and best produced results (outputs). The model configuration for each test consisted of a Window size, Training Sample size, Validation Sample size, Filter size, Kernel size, Batch size and Epoch size. Each of the 5 experiments were performed on a computer with specifications: Intel i7 processor, Nvidia GTX 1080 and 16GB Ram.

### A. Experiment 1 and Results

TABLE VI. W=500

| Inputted Configuration | | Outputted Results | |
| --- | --- | --- | --- |
| Window Size | 500 | Accuracy | 0.9377% |
| Training Samples | 108276 | Loss | 0.1641% |
| Validation Samples | 12031 | Validation Accuracy | 0.9403% |
| n_Filters | 150 | Validation Loss | 0.1640% |
| k_Size | 150 | Sensitivity (Recall) | 0.9705 |
| Batch_Size | 8192 | Specificity | 0.9725 |
| Epochs | 50 | F1_Score | 0.9717 |
| --- | | Kappa_Score | 0.9430 |
| --- | | Log_Loss | 0.0836 |
| --- | | ROCAUC | 0.9945 |

Table VI, 'Inputted Configuration' column presents the optimal configuration for dataset W=500. For this model, using a maximum workable batch size of 8192, a kernel size of 150 and 150 filters, was empirically found to return the best results when run over a period of 50 epochs. The overall performance of this model, presented in Table VI 'Outputted Results' column, is very satisfying, particularly when based on the high scores of Sensitivity and Specificity. This was found to be our best performing model.

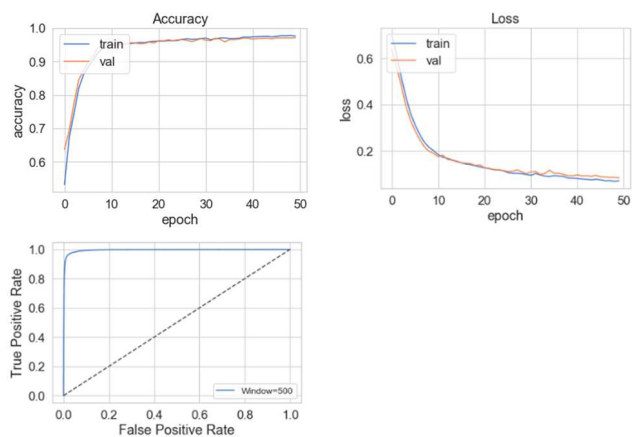

Fig. 2. Graphical output results from our 1DCNN model using dataset W=500. It shows Training and Validation accuracy, Training and Validation Loss, and ROCAUC.

Fig. 2 presents training and validation accuracy, logloss and ROCAUC plots for Table VI results. Looking at both the training and validation accuracy and logloss plots, it is possible to see early convergence of the values at approx. 0.9 (accuracy) and 0.2 (loss), and at approx. 10 epochs. The plots also show the model is still slightly improving over time, with almost smooth and flattening lines. This indicates the model is working well and no signs of overfitting. More evidence to the excellence of this model is demonstrated by the ROCAUC graph. The ROC curve is very close to the top left-hand corner and the AUC is very almost at its highest limit of 1. These two values indicate this model has a very high accuracy when predicting between the two classes.

### B. Experiment 2 and Results

TABLE VII.    W=1000

| Inputted Configuration | | Outputted Results | |
|---|---|---|---|
| Window Size | 1000 | Accuracy | 0.9528% |
| Training Samples | 54136 | Loss | 0.1364% |
| Validation Samples | 6016 | Validation Accuracy | 0.9505% |
| n_Filters | 250 | Validation Loss | 0.1433% |
| k_Size | 250 | Sensitivity (Recall) | 0.9612 |
| Batch_Size | 4096 | Specificity | 0.9730 |
| Epochs | 50 | F1_Score | 0.9669 |
| --- | | Kappa_Score | 0.9342 |
| --- | | Log_Loss | 0.0998 |
| --- | | ROCAUC | 0.9935 |

In Table VII, 'Inputted Configuration' shows the optimal configuration for Dataset W=1000, which includes a batch-size (4096) half of that to the previous W=500 model, but a filter number (250) and a kernel size (250) of almost double. These inputted parameters were empirically found to return the best results for this model when averaged over 50 epochs. Overall the 'Outputted Results' shows the performance of the model were very similar to that of the W=500 model. The training and validation accuracy and logloss were slightly higher than W=500, but the more significant measurements of Sensitivity and Specificity were lower.

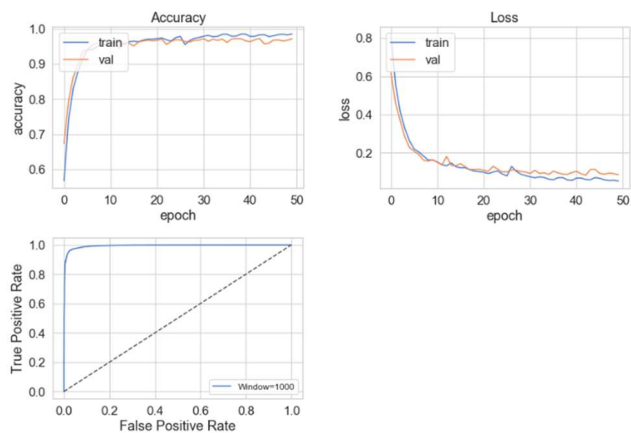

Fig. 3. Graphical output results from our 1DCNN model using dataset W=1000. It shows Training and Validation accuracy, Training and Validation Loss, and ROCAUC.

Again, almost identical to the previous model, both the training and validation accuracy and logloss plots in Fig. 3, show the values converging at approximately 0.9 (accuracy) and 0.2 (loss), and at approx. 10 epochs. However, although it is clear to see the model is still improving at around 50 epochs, it is also possible to see both values highly fluctuating, which indicate this model doesn't seem to be operating as seamless as the previous model. Yet, this is still classed as an excellent model with a high level of class separation, as demonstrated by the ROCAUC graph results.

### C. Experiment 3 and Results

TABLE VIII.    W=1500

| Inputted Configuration | | Outputted Results | |
|---|---|---|---|
| Window Size | 1500 | Accuracy | 0.9161% |
| Training Samples | 36090 | Loss | 0.2131% |
| Validation Samples | 4010 | Validation Accuracy | 0.9095% |
| n_Filters | 100 | Validation Loss | 0.2312% |
| k_Size | 1000 | Sensitivity (Recall) | 0.9592 |
| Batch_Size | 4096 | Specificity | 0.9472 |
| Epochs | 50 | F1_Score | 0.9536 |
| --- | | Kappa_Score | 0.9064 |
| --- | | Log_Loss | 0.1374 |
| --- | | ROCAUC | 0.9861 |

The inputted parameters for the W=1500 model, shown in Table VIII, required a much greater kernel size (1000) and a reduction in filters (100) to that of the previous models. The overall outputted results produced by this model show that it didn't outperform the previous two models (W500, W1000), however, the results are still classed as very good.

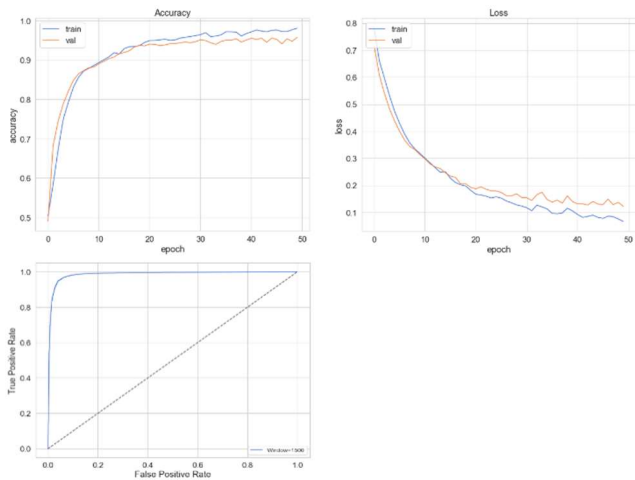

Fig. 4. Graphical output results from our 1DCNN model using dataset W=1500. It shows Training and Validation accuracy, Training and Validation Loss, and ROCAUC

Training and validation accuracy and logloss plots in Fig. 4, show improvement over time, however validation is at a slower rate and causing the model to start overfitting at approx. 20 epochs. The addition of a dropout layer or stopping the training early could solve this issue.

### D. Experiment 4 and Results

TABLE IX.  W=2000

| Inputted Configuration | | Outputted Results | |
|---|---|---|---|
| Window Size | 2000 | Accuracy | 0.9086% |
| Training Samples | 27065 | Loss | 0.2754% |
| Validation Samples | 3008 | Validation Accuracy | 0.9011% |
| n_Filters | 100 | Validation Loss | 0.2893% |
| k_Size | 500 | Sensitivity (Recall) | 0.9575 |
| Batch_Size | 4096 | Specificity | 0.9702 |
| Epochs | 50 | F1_Score | 0.9634 |
| --- | | Kappa_Score | 0.8959 |
| --- | | Log_Loss | 0.1571 |
| --- | | ROCAUC | 0.9855 |

In Table IX, the W=2000 model used an almost identical set of parameters to the previous model W=1500, but with a much kernel length half the size. These parameters were empirically found to provide the best results for this model. The overall outputted results from this model were very good, particularly when examining the high scores produced by sensitivity and specificity, which were almost equal to our best performing model W=500.

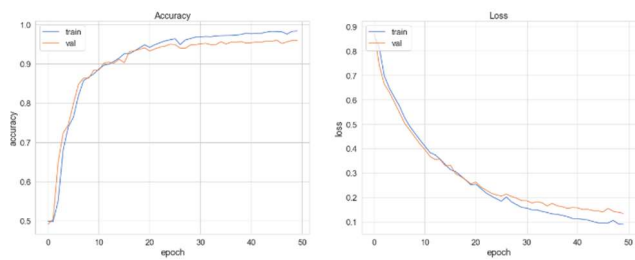

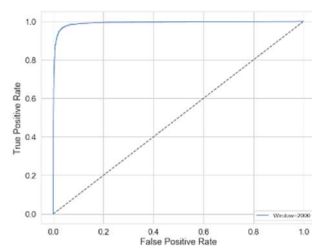

Fig. 5. Graphical output results from our 1DCNN model using dataset W=2000. It shows Training and Validation accuracy, Training and Validation Loss, and ROCAUC.

At 20 epochs the training and validation accuracy and logloss plots in Fig.5 shows an improved learning curve, however, similar to the previous model the validation lines starts to drift and begins to display signs of overfitting. Again, ending the training early, at around 20 epochs will solve this. The ROCAUC also indicated the model has a high accuracy predicting the presence and absence of Apnoea.

### E. Experiment 5 and Results

TABLE X.  W=2500

| Inputted Configuration | | Outputted Results | |
|---|---|---|---|
| Window Size | 2500 | Accuracy | 0.9046% |
| Training Samples | 21643 | Loss | 0.2605% |
| Validation Samples | 2405 | Validation Accuracy | 0.9067% |
| n_Filters | 100 | Validation Loss | 0.2760% |
| k_Size | 800 | Sensitivity (Recall) | 0.9414 |
| Batch_Size | 4096 | Specificity | 0.9545 |
| Epochs | 50 | F1_Score | 0.9479 |
| --- | | Kappa_Score | 0.8959 |
| --- | | Log_Loss | 0.1571 |
| --- | | ROCAUC | 0.9855 |

The final experiment was performed with W=2500 (Table X), which used similar input parameters to the previous two models (W=1500, W=2000) and included a batch file of 4096, a high kernel size (800), a 100 filters and averaged over 50 epochs. Again, these parameters were empirically found to return the best results for this model.

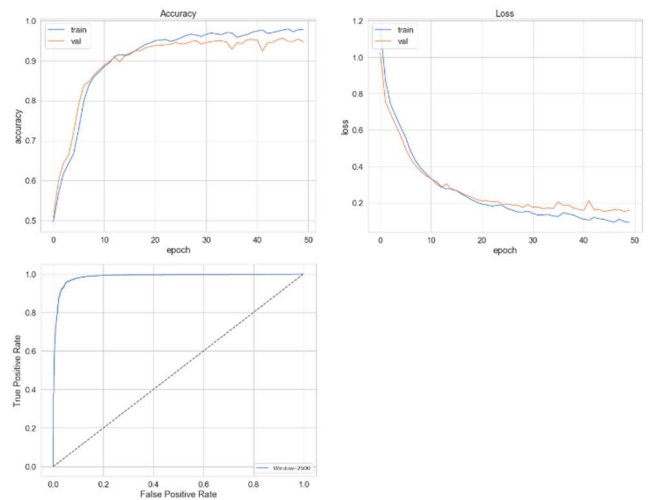

Fig. 6. Graphical output results from our 1DCNN model using dataset W=2500. It shows Training and Validation accuracy, Training and Validation Loss, and ROCAUC.

The training and validation accuracy and logloss plots for this model, seen in Fig.6, are very good and almost identical to those produced by the W=2000 model. They show a strong sense of learning, albeit, with small degree of fluctuation before the values start to flatten at around 50 epochs. The ROCAUC graph is also very good. Akin to all the ROCAUC graphs produced across all experiments, it shows the model is very good at distinguishing between the two classes.

IV. DISCUSSION AND CONCLUSION

Obstructed Sleep Apnoea is a debilitating condition that can lead to mortality and other serious complications. It is reported that up to 20% of the general population suffer with OSA and over 80% of patients remain incorrectly diagnosed. Traditional diagnostic techniques for OSA are expensive, cumbersome, complex and lengthy, meaning sufferers do not receive the required treatment and therapy in adequate time. In an attempt to combat these issues, many studies took a computerised approach through the application of machine learning methods. However, these approaches require solid domain knowledge and extensive manual input, thus they can be problematic, lengthy and expensive when attempting to produce high-end results. Furthermore, such methods are not always successful, and their prediction efficiency has been called into question.

Our solution addresses many of these issues. We propose a novel method to automatically detect OSA through single lead ECG data using a CNN model. By acquiring trusted OSA signal data, evaluated and selected through its extensive use in previous high-quality studies. We uniquely created 5 balanced datasets of different window sizes. Since the nature of the captured data was time series data, the proposal of a 1DCNN was deemed as the best model for data classification. The architecture of the model was designed to mathematically calculate and produce maximum results in minimum time by using specific mathematical mechanisms consisting of single kernel, 1 Convolutional layer, 1 Max Pooling layer and a Fully Connected Multilayer Perceptron (MLP) consisting of 1 Hidden layer and a Softmax output. Empirically, one convolutional layer provided us with the best produced results. Adding additional layers didn't improve the model.

The robustness and effectiveness of our proposed model was evaluated through an experiment designed to test and train the model. This was conducted by running our 1DCNN over the 5 uniquely developed datasets with large sample sizes, each time setting varying depths and values of configuration parameters, n_filters, k_sizes, batch sizes and epochs. To find the model's optimal performance, we ran this process over 1200 times (approx. 300 per dataset window). Evaluation and scoring were achieved using a series of standard performance metrics outlined in section *(G) Performance Metrics*. All our tests were run with peak and top-end batch sizes, as well as large kernel sizes, ensuring our model was producing the truest accuracy possible.

At this stage, our CNN model is demonstrating excellent ability in identifying the presence and absence of apnoea for both new and unseen data. This is quite evident when observing the metric results and associated graphs across all 5 windowing experiments (Tables VI – X, Figs. 2 – 6), particularly the high classification measurements of sensitivity and specificity and the very good optimised performance results produced by Training and Validation (Accuracy and Loss). Studying the graphs produced by our best performing model (W=500) in experiment 1, it is clear to see both value lines (Training and Validation) begin to converge very early in the test before smoothing and almost flattening, with no sign of overfitting. To see if we could further improve the models learning, we ran longer tests at 100, 150 and 200 epochs, however, it soon became apparent that the model reaches its optimum performance and starts to flatline between 10 to 20 epochs and approx. 25 to 40 seconds.

So far, our experiments have provided us with very satisfying results and show how a 1DCNN model can help to benefit many of the current issues that hinder the swift identification of Obstructed Sleep Apnoea. Future testing will be performed using data collected from our own subjects. The results of this study will be implemented and published in our future papers.

ACKNOWLEDGMENTS

The dataset for this study was sourced from Physiobank, which is a subdivision of the publicly accessible and well renowned on-line data exchange site, Physionet. PhysioNet is a web-based library of physiological data, accompanied by analytic software and is supported by the National Institute of General Medical Sciences (NIGMS) and the National Institute of Biomedical Imaging and Bioengineering (NIBIB) under NIH grant number 2R01GM104987-09. https://archive.physionet.org/physiobank/. The authors would like to thank all those involved in making the dataset available to the general public.

REFERENCES

[1] D. P. White, 'Sleep-related breathing disorder.2. Pathophysiology of obstructive Sleep Apnoea.', *Thorax*, vol. 50, no. 7, pp. 797–804, 1995.

[2] B. S. Young T, Palta M, Dempsey J, Skatrud J, Weber S, 'The occurrence of sleep-disordered breathing among middle-aged adults', *N Engl J Med 1993;3281230-5.*, vol. 3, no. 3, p. 4, 1993.

[3] R. Cartwright, 'Obstructive sleep apnea: A sleep disorder with major effects on health', *Disease-a-Month*, vol. 47, no. 4, pp. 105–147, 2001.

[4] S. M. Ejaz, I. S. Khawaja, S. Bhatia, and T. D. Hurwitz, 'Obstructive sleep apnea and depression: A review', *Innov. Clin. Neurosci.*, vol. 8, no. 8, pp. 17–25, 2011.

[5] T. Young, L. Evans, L. Finn, and M. Palta, 'Estimation of the clinically diagnosed proportion of sleep apnea syndrome in middle-aged men and women', *Sleep*, vol. 20, no. 9, pp. 705–706, 1997.

[6] Frost & Sullivan, 'Hidden Health Crisis Costing America Billions: Underdiagnosing and Undertreating Obstructive Sleep Apnea Draining Healthcare System', *Am. Acad. Sleep Med. 2016.*, 2016.

[7] N. M. Punjabi *et al.*, 'Sleep-disordered breathing and mortality: A prospective cohort study', *PLoS Med.*, vol. 6, no. 8, 2009.

[8] T. Young, 'Epidemiological insights into the public health burden of sleep disordered breathing: Sex differences in survival among sleep clinic patients', *Thorax*, vol. 53, no. SUPPL. 3, pp. 16–19, 1998.

[9] J. Teran-Santos, A. Jimenez-Gomez, J. Cordero-Guevara, 'Association Between Sleep Apnea and the Risk of Traffic Accidents', *N Engl J Med*, vol. 340, pp. 847–851, 1999.

[10] W. T. McNicholas, 'Diagnosis of obstructive sleep apnea in adults', *Proc. Am. Thorac. Soc.*, vol. 5, no. 2, pp. 154–160, 2008.

[11] F. Chung, H. R. Abdullah, and P. Liao, 'STOP-bang questionnaire a practical approach to screen for obstructive sleep apnea', *Chest*, vol. 149, no. 3, pp. 631–638, 2016.

[12] N. C. Netzer, R. A. Stoohs, C. M. Netzer, K. Clark, and K. P. Strohl, 'Using the Berlin Questionnaire to identify patients at risk for the sleep apnea syndrome', *Annals of Internal Medicine*, vol.

131, no. 7. pp. 485–491, 1999.

[13] M. W. Johns, 'A new method for measuring daytime sleepiness: the Epworth sleepiness scale', *Sleep*, vol. 14, no. 6, pp. 540–545, 1991.

[14] A. L. Chesson *et al.*, 'The indications for polysomnography and realted procedures. An American sleep disorders association review', *Pneumologie*, vol. 52, no. 3, p. 154, 1998.

[15] W. W. Flemons, N. J. Douglas, S. T. Kuna, D. O. Rodenstein, and J. Wheatley, 'Access to Diagnosis and Treatment of Patients with Suspected Sleep Apnea', *Am. J. Respir. Crit. Care Med.*, vol. 169, no. 6, pp. 668–672, 2004.

[16] N. A. Collop *et al.*, 'Clinical guidelines for the use of unattended portable monitors in the diagnosis of obstructive sleep apnea in adult patients. Portable Monitoring Task Force of the American Academy of Sleep Medicine.', *J. Clin. Sleep Med.*, vol. 3, no. 7, pp. 737–47, 2007.

[17] L. Abrahamyan *et al.*, 'Diagnostic accuracy of level IV portable sleep monitors versus polysomnography for obstructive sleep apnea: a systematic review and meta-analysis', *Sleep Breath.*, vol. 22, no. 3, pp. 593–611, 2018.

[18] V. K. Kapur *et al.*, 'Clinical practice guideline for diagnostic testing for adult obstructive sleep apnea: An American academy of sleep medicine clinical practice guideline', *J. Clin. Sleep Med.*, vol. 13, no. 3, pp. 479–504, 2017.

[19] L. Almazaydeh, K. Elleithy, and M. Faezipour, 'Detection of obstructive sleep apnea through ECG signal features', *IEEE Int. Conf. Electro Inf. Technol.*, pp. 1–6, 2012.

[20] S. G. Jones *et al.*, 'Regional Reductions in Sleep Electroencephalography Power in Obstructive Sleep Apnea: A High-Density EEG Study', *Sleep*, vol. 37, no. 2, pp. 399–407, 2014.

[21] J. V. Marcos, R. Hornero, D. Álvarez, M. Aboy, and F. Del Campo, 'Automated prediction of the apnea-hypopnea index from nocturnal oximetry recordings', *IEEE Trans. Biomed. Eng.*, vol. 59, no. 1, pp. 141–149, 2012.

[22] N. Selvaraj and R. Narasimhan, 'Detection of sleep apnea on a per-second basis using respiratory signals', *Proc. Annu. Int. Conf. IEEE Eng. Med. Biol. Soc. EMBS*, pp. 2124–2127, 2013.

[23] B. L. Koley and D. Dey, 'Automatic detection of sleep apnea and hypopnea events from single channel measurement of respiration signal employing ensemble binary SVM classifiers', *Meas. J. Int. Meas. Confed.*, vol. 46, no. 7, pp. 2082–2092, 2013.

[24] A. Azarbarzin and Z. Moussavi, 'Snoring sounds variability as a signature of obstructive sleep apnea', *Med. Eng. Phys.*, vol. 35, no. 4, pp. 479–485, 2013.

[25] E. Dafna, A. Tarasiuk, and Y. Zigel, 'Automatic detection of whole night snoring events using non-contact microphone', *PLoS One*, vol. 8, no. 12, 2013.

[26] J. Solà-Soler, J. A. Fiz, J. Morera, and R. Jané, 'Multiclass classification of subjects with Sleep Apnoea-hypopnoea syndrome through snoring analysis', *Med. Eng. Phys.*, vol. 34, no. 9, pp. 1213–1220, 2012.

[27] T. Penzel, G. B. Moody, R. G. Mark, A. L. Goldberger, and J. H. Peter, 'Apnea-ECG database', *Comput. Cardiol.*, pp. 255–258, 2000.